\documentclass[a4paper]{JHEP3}

\usepackage{epsfig}
\usepackage{amsmath}

\title{On the existence of a static black hole on a brane}

\author{Hirotaka Yoshino\\
Department of Physics, University of Alberta, 
Edmonton, Alberta, Canada T6G 2G7
E-mail: \email{hyoshino@phys.ualberta.ca}
}

\preprint{Alberta-Thy-20-08}

\abstract{
We study a static black hole 
localized on a brane in the Randall-Sundrum (RS) II
braneworld scenario. To solve this problem numerically,
we develop a code  
having the almost 4th-order accuracy.
This code derives the highly accurate result 
for the case where the brane tension is zero, i.e., 
the spherically symmetric case. 
However, a nonsystematic error is detected
in the cases where the brane tension is nonzero.
This error is irremovable by any systematic methods
such as increasing the resolution, setting the 
outer boundary at more distant location, or improving
the convergence of the numerical relaxation. 
We discuss the possible origins
for the nonsystematic error, and conclude that 
our result is naturally interpreted as
the evidence for the nonexistence of solutions
to this setup, 
although an ``approximate'' solution exists
for sufficiently small brane tension. 
We discuss the possibility that the black holes produced 
on a brane may be
unstable and lead to two interesting consequences:
the event horizon pinch and the brane pinch.
 }

\begin{document}

%
%======================================%
%<<<<<<<<<<<< SECTION I  >>>>>>>>>>>>>>%
%======================================%
%
\section{Introduction}
\label{Sec:I}

The Randall-Sundrum (RS) braneworld scenarios \cite{RS99-1,RS99-2}
have been attracting a lot of attentions
since their appearance in 1999.
In these scenarios, our world is a 5-dimensional
anti--de-Sitter (AdS) spacetime
with a negative cosmological constant $\Lambda$.
In the RS I scenario \cite{RS99-1}, there are two branes with positive
and negative tension, and our 3-dimensional space is the negative
tension brane. In this scenario, the Planck energy
could be $O(\mathrm{TeV})$ and the black hole production
might happen at the Large Hadron Collider (LHC) \cite{BF99,DL01,GT02}.
In the RS II scenario \cite{RS99-2}, there is only one brane
with positive tension and the extra-dimension
is infinitely extended. By the bulk curvature effect,
the 4-dimensional gravity is successfully realized
on the brane~\cite{GT99,GKR00}. Although the
hierarchy problem is not solved, this scenario
is also interesting since it 
gives the new mechanism of the compactification
of the extra dimension.

In this paper, we study a static black hole
localized on the brane in the RS II scenario. 
There are at least three motivations for this study.
The first motivation is related to the astrophysics in this scenario.
The final state of the gravitational collapse
of a star is expected to be different from the
4-dimensional black hole, since the event horizon
extends into the bulk. By studying black hole solutions,
we might obtain observable features of braneworld black holes
or constraints for the RS models. 
The second motivation is the black hole production at the LHC 
\cite{BF99,DL01,GT02}.
Although the study of the black holes in the RS II scenario
is not directly related to this phenomena, the obtained
results can be translated into the RS I case by changing the
sign of the bulk curvature scale $\ell$ if the effect of the
existence of the positive tension brane is neglected.
Therefore, we would obtain some indication for the effect
of the bulk curvature on the black hole physics that might
be observed at the LHC.

The third motivation is the AdS/CFT correspondence,
which conjectures that the gravitational theory in the 
AdS spacetime is dual to the conformal field theory (CFT)
on the boundary of that spacetime. It was suggested that
the AdS/CFT correspondence hold also for the RS II scenario
\cite{HHR00,DL00,SI01}.
Based on the AdS/CFT correspondence,
it was conjectured that no solution of a large black hole
on a brane exists \cite{T02,EFK02}. 
Their argument is that a 4-dimensional black hole
with quantum fields is dual to a classical 5-dimensional black hole
on a brane,
and the dual phenomena of evaporation of a 4-dimensional
black hole is expected to be escape of a 5-dimensional black hole
into the bulk. 
However, there are different opinions on how to apply
the AdS/CFT correspondence and thus on the expected black hole 
solution space \cite{ABF04,FFNOS05,FRW06,FP07-1,T07,GRZ08}. 
The explicit study of the black hole solution
would shed a new light on this issue.

There were many attempts to obtain solutions of
a black hole localized on a brane. 
The exact solution of a static black hole intersecting 
a 2-brane in a 4-dimensional spacetime was 
found in \cite{EHM99,EHM99-2} using the so-called C-metric \cite{PD76}.
The brane is given as some (2+1)-dimensional slice
in the spacetime of the C-metric, 
and the geometry of the brane is asymptotically a cone. 
In contrast to the 4-dimensional case, finding solutions of a
black hole on a brane turned out to be very difficult 
in the 5-dimensional case. For several years,
no successful discovery of the solution was not 
reported, although several efforts were made
\cite{DMPR00,CRSS00,KT01,KOT03,CFM02,CM03,CR03}.
%Some authors \cite{BGM01} even stated that there might not exist
%solutions of a static black hole on a brane in the 5-dimensional case.

After that, Kudoh {\it et al.}~\cite{KTN03}
reported a numerical study of a static black hole on a brane.
In their formulation, the problem was reduced to 
elliptic equations for metric functions with appropriate
boundary conditions. They solved this problem by a relaxation method.
For the cases where the horizon radius is much smaller than
the bulk curvature scale ($L:=\ell/\rho_h\gg 1$),
they realized the sufficient convergence. As the mass of the black hole
becomes large, the convergence became worse and the error grew.
For this reason, they only showed the results for $3\le L\le 500$.
Since their study, no calculation has been performed for
the case $L\lesssim 1$. The result of \cite{KTN03} might be interpreted as
an evidence for the existence of solutions of a black hole on a brane
with a small mass.
However, the growth of the error with $(1/L)$ also can be regarded 
as the evidence for nonexistence of such solutions. 
Therefore, we have to be careful in interpreting their results,
and this study has to be reexamined.

The solution of a static black hole on a brane
was studied also by the perturbative method
by Karasik {\it et al.} \cite{KSSW03,KSSW04}.
They studied the situation where the black hole radius
is much smaller than the bulk curvature scale using the
matching method, and derived the solution that is
at least $C^1$ on the horizon. In Appendix A,
we summarize the part of their results that 
is closely related to the discussions in this paper.
As pointed out there (and also in the original paper \cite{KSSW04}),
there remains a possibility that the horizon becomes singular
at the higher-order perturbations. Therefore, this study also cannot
be regarded as the rigorous evidence for the existence of the
solution. 
Another related study is the perturbative study of a 
higher-dimensional version of the C-metric by Kodama \cite{Kodama08}.
Using the gauge invariant formulation for the perturbation
of a higher-dimensional Schwarzschild black hole \cite{KI03,IK03},
he succeeded in deriving the solution of a 5-dimensional black hole 
accelerated by a string with uniform tension.
However, in contrast to the 4-dimensional case, there was no
(3+1)-dimensional slice that realizes the junction condition on the brane in 
that spacetime.

As summarized above, there is no rigorous answer to the solution space
of a static black hole on a brane. We have to examine the (non)existence
of the solution, and if it exists, the
range of $L\lesssim 1$ has to be explored.
Although we basically follow the formulation
by Kudoh {\it et al.}~\cite{KTN03},
a significantly improved numerical method is developed. 
Specifically, we focus our attention to the error analysis
taking account of the possibility that no solution exists. 
It is usually said that numerical calculations cannot
prove nonexistence of solutions rigorously.
However, since our code has the sufficient accuracy, 
it is possible to obtain a strong indication for 
(non)existence of the solution
by such an error analysis. In this way,
we obtain the result that at least raises strong doubts 
for the existence of solutions of a black hole on a brane.

This paper is organized as follows.
In Sec.~\ref{Sec:II}, we explain the setup 
of the problem, which is a short review of the corresponding
part of Ref.~\cite{KTN03}.
In Sec.~\ref{Sec:III}, our numerical method is explained
taking attention to the difference from the code of Ref.~\cite{KTN03}.
In Sec.~\ref{Sec:IV}, we show the results for the case where
the brane tension is zero, i.e. the spherically symmetric case.
This proves the correctness and high accuracy of our code.
After discussing the features that the numerical solutions
have to satisfy, 
we introduce {\it ``systematic error''}
and {\it ``nonsystematic error''}, which are the key notions
in order to interpret our results for the cases
where the brane tension is nonzero.  
In Sec.~\ref{Sec:V}, the numerical results for the cases
of a nonzero tension brane are shown taking attention
to the numerical errors, and the existence of the nonsystematic error
is proved. We also point out that our results are consistent with the
perturbative study \cite{KSSW03,KSSW04} 
if the numerical errors are ignored.
Then, we discuss the interpretation of our results.
Our results are naturally interpreted 
as the evidence for the nonexistence of solutions
of a black hole on a brane.
Section~\ref{Sec:VI} is devoted to summary and discussion.
Assuming the nonexistence of solutions of a static 
black hole on a brane, we discuss possible phenomena
that could happen after the black hole formation 
on a brane.
In Appendix \ref{Sec:A}, we briefly summarize the results of the perturbative
study \cite{KSSW03,KSSW04} of a small black hole on a brane, taking
attention to the part that is closely related to this paper.

%
%======================================%
%<<<<<<<<<<<< SECTION II  >>>>>>>>>>>>>>%
%======================================%
%
\section{Setup}
\label{Sec:II}

In this section, we explain the setup to study
a static black hole localized on a brane, 
which is spherically symmetric
on the brane (in the 4-dimensional sense) and 
axisymmetric in the bulk spacetime (in the 5-dimensional sense).
Since we basically follow the formulation of Kudoh {\it et al.},
this is a brief review of the corresponding part of Ref.~\cite{KTN03}.

We assume the following metric ansatz:
%===========<Equation>============%
%
\begin{equation}
ds^2=\frac{\ell^2}{z^2}
\left[
-T^2dt^2+e^{2R}(dr^2+dz^2)+r^2e^{2C}d\Omega_2^2
\right],
\label{original-metric}
\end{equation}
%
%=================================%
where $d\Omega_2^2:=d\theta^2+\sin^2\theta d\phi^2$. Here, 
$\ell$ is related to the bulk
cosmological constant as $\Lambda=-6/\ell^2$. 
The functions $T$, $R$, and $C$ depend only on $z$ and $r$.
Each hypersurface $z=\textrm{const.}$ is
a 4-dimensional spherically symmetric spacetime
where $r$ is the radial coordinate. 
$r=0$ is the symmetry axis in the 5-dimensional
point of view. If we set 
$T=1$ and $R=C=0$, the spacetime is the AdS spacetime
in the Poincar\'e coordinates. 
In the metric~\eqref{original-metric}, 
the fact that
the timelike killing vector is hypersurface orthogonal
is used following the
standard definition of a static spacetime
(e.g. Ref.~\cite{W84}).
The fact 
that a 2-dimensional surface is conformally
flat is also used in \eqref{original-metric}. 
For these reasons, the assumed form of the metric \eqref{original-metric}
is sufficiently general for our setup.

We introduce the coordinates $\rho$ and $\chi$
by
%===========<Equation>============%
%
\begin{eqnarray}
r=\rho\sin\chi; \qquad
z=\ell+\rho\cos\chi.
\end{eqnarray}
%
%=================================%
Then, $T$, $R$, and $C$
are the functions of  $\rho$ and $\chi$.
The locations of the event horizon and the brane are assumed
to be $\rho=\rho_h$ and $\chi=\pi/2$, respectively.
Note that it was shown in \cite{KTN03} that 
this assumption is possible in general
using the remaining gauge degree of 
freedom in the metric. 
Although there are the two parameters $\ell$ and $\rho_h$
in this system, the spacetime
is basically specified by the dimensionless parameter 
%===========<Equation>============%
%
\begin{equation}
L:=\ell/\rho_h,
\end{equation}
%
%=================================%
since fixing $L$ and changing $\ell$ or $\rho_h$ corresponds
to just changing the unit of the length.
In the numerical calculation,  
we put the event horizon at 
$\rho_h=1$. In this choice of the length unit, we have $L=\ell$.
In this paper, we basically use $(1/L)$ as the parameter
to specify the spacetime rather than $L$.

%
%======================================%
%<<<<<<<<<< subsection 2.1 >>>>>>>>>>>>%
%======================================%
%
\subsection{Equations}
\label{Sec:IIA}

The equations to be solved 
are the Einstein equations of a vacuum spacetime
with a negative cosmological constant,
%===========<Equation>============%
%
\begin{equation}
\mathcal{G}_{\mu\nu}:=R_{\mu\nu}-(2/3)\Lambda g_{\mu\nu}=0.
\end{equation}
%
%=================================%
From $\mathcal{G}_{t}^{t}=0$, 
$\mathcal{G}_{t}^{t}-\mathcal{G}_{\rho}^{\rho}
-\mathcal{G}_{\chi}^{\chi}+2\mathcal{G}_{\theta}^{\theta}=0$ 
and $\mathcal{G}_{\theta}^{\theta}=0$,
the elliptic equations for $T$, $R$, and $C$ are obtained:
%===========<Equation>============%
%
\begin{multline}
\nabla^2T
+2\left(C_{,\rho}+\frac{2\ell}{z\rho}-\frac{1}{\rho}\right)T_{,\rho}
+\frac{2}{\rho^2}\left(\cot\chi+C{,\chi}+\frac{2\rho}{z}\sin\chi\right)T_{,\chi}
\\
+\frac{2}{\rho z}\left(\sin\chi C_{,\chi}-\cos\chi\rho C_{,\rho}\right)T
+\frac{4}{z^2}\left(1+\frac{\Lambda \ell^2}{6}e^{2R}\right)T=0;
\label{EE1}
\end{multline}
%
%=================================%
%===========<Equation>============%
%
\begin{multline}
\nabla^2 R-\frac{1-e^{2(R-C)}}{\rho^2\sin^2\chi}
-\frac{2}{z^2}\left(1+\frac{\Lambda \ell^2}{6}e^{2R}\right)
-\frac{2T_{,\rho}}{T}\left(C_{,\rho}+\frac{\ell}{z\rho}\right)
-\frac{2T_{,\chi}}{\rho^2 T}\left(\cot\chi+C_{,\chi}+\frac{\rho}{z}\sin\chi\right)
\\
-C_{,\rho}\left(C_{,\rho}+\frac{4\ell}{z\rho}-\frac{2}{\rho}\right)
-\frac{C_{,\chi}}{\rho^2}\left(C_{,\chi}+2\cot\chi+\frac{4\rho}{z}\sin\chi\right)=0;
\label{EE2}
\end{multline}
%
%=================================%
%===========<Equation>============%
%
\begin{multline}
\nabla^2 C+\frac{1-e^{2(R-C)}}{\rho^2\sin^2\chi}
+\frac{4}{z^2}\left(1+\frac{\Lambda \ell^2}{6}e^{2R}\right)
+\frac{T_{,\rho}}{T}\left(C_{,\rho}+\frac{\ell}{z\rho}\right)
+\frac{T_{,\chi}}{\rho^2 T}\left(\cot\chi+C_{,\chi}+\frac{\rho}{z}\sin\chi\right)
\\
+C_{,\rho}\left(2C_{,\rho}+\frac{5\ell}{z\rho}-\frac{1}{\rho}\right)
+\frac{C_{,\chi}}{\rho^2}\left(2C_{,\chi}+4\cot\chi+\frac{5\rho}{z}\sin\chi\right)=0.
\label{EE3}
\end{multline}
%
%=================================%
Here, $\nabla^2:=\partial_\rho^2+\partial_\rho/\rho+\partial_\chi^2/\rho^2$.
There are two other equations 
derived by $\mathcal{G}_{\rho\theta}=0$ and 
$\mathcal{G}_{t}^{t}-\mathcal{G}_{\rho}^{\rho}
+\mathcal{G}_{\chi}^{\chi}+2\mathcal{G}_{\theta}^{\theta}=0$,
respectively:
%===========<Equation>============%
%
\begin{multline}
\frac{1}{2T}
\left[T_{,\rho\chi}-R_{,\chi}T_{,\rho}-T_{,\chi}\left(R_{,\rho}+\frac{1}{\rho}\right)\right]
+C_{,\rho\chi}+C_{,\rho}(\cot\chi+C_{,\chi})
\\
+\frac{R_{,\chi}}{2}\left(\frac{1}{\rho}-2C_{,\rho}-\frac{3\ell}{z\rho}\right)
-R_{,\rho}\left(C_{,\chi}+\cot\chi+\frac{3\rho}{2z}\sin\chi\right)
=0;
\label{EE4}
\end{multline}
%
%=================================%
%===========<Equation>============%
%
\begin{multline}
\frac{1}{T}
\left[T_{,\chi\chi}
+T_{,\chi}\left(2\cot\chi+2C_{,\chi}-R_{,\chi}+\frac{3\rho}{z}\sin\chi\right)
+\rho T_{,\rho}\left(2\rho C_{,\rho}+\rho R_{,\rho}+\frac{3\ell}{z}\right)
\right]
\\
+2C_{,\chi\chi}+C_{,\chi}\left(6\cot\chi+3C_{,\chi}+\frac{6\rho}{z}\sin\chi\right)
+\rho C_{,\rho}\left(2\rho R_{,\rho}+\rho C_{,\rho}-2+\frac{6\ell}{z}\right)
\\
-R_{,\chi}\left(2\cot\chi+\frac{3\rho}{z}\sin\chi+2C_{,\chi}\right)
+\rho R_{,\rho}\left(\frac{3\ell}{z}-1\right)
\\
+\frac{1-e^{2(R-C)}}{\sin^2\chi}
+\frac{6\rho^2}{z^2}\left(1+\frac{\Lambda\ell^2}{6}e^{2R}\right)=0.
\label{EE5}
\end{multline}
%
%=================================%
It was shown in \cite{KTN03} that 
a solution of Eqs.~\eqref{EE1}--\eqref{EE3} 
automatically satisfies Eqs.~\eqref{EE4} and \eqref{EE5} provided that
Eq.~\eqref{EE4} is satisfied on the boundary and Eq.~\eqref{EE5} 
is satisfied at one point.
Therefore, the latter two equations \eqref{EE4} and \eqref{EE5}
are the constraint equations, and the main equations 
are the first three equations \eqref{EE1}--\eqref{EE3}.

In the limit $(1/L):=\rho_h/\ell\to 0$,  
the solution is 
%===========<Equation>============%
%
\begin{equation}
T_0=\frac{\rho^2-1}{\rho^2+1}; \qquad
R_0=C_0=\log\left(1+\frac{1}{\rho^2}\right),
\label{spherical-case}
\end{equation}
%
%=================================%
which gives the metric of 
the Schwarzschild black hole in the isotropic coordinates.
This fact raised the expectation 
that the Schwarzschild-like black hole solution exists for 
$(1/L)\ll 1$ 
in the previous studies \cite{KTN03,KSSW03,KSSW04}.

%
%======================================%
%<<<<<<<<<< subsection 2.2 >>>>>>>>>>>>%
%======================================%
%
\subsection{Boundary conditions}
\label{Sec:IIB}

There are four boundaries of the computation domain:
the outer boundary $\rho=\rho_{\rm out}$, the brane $\chi=\pi/2$,
the symmetry axis $\chi=0$, and the horizon $\rho=\rho_h$.
Here we summarize the boundary conditions for $T$, $R$, and $C$
on the four boundaries one by one.

At the distant region, the spacetime should be reduced to the AdS spacetime.
Therefore, we impose
%===========<Equation>============%
%
\begin{equation}
T=1;\qquad C=0; \qquad R=0,
\label{outer-boundary}
\end{equation}
%
%=================================%
at $\rho=\rho_{\rm out}$ for a sufficiently large $\rho_{\rm out}$.
Strictly speaking, we have to impose the conditions \eqref{outer-boundary}
at $\rho=\infty$. 
Since we adopt the finite value of $\rho_{\rm out}$
for numerical convenience, we have to keep in mind 
the existence of an error from the finiteness of $\rho_{\rm out}$
(say, the {\it ``boundary error''})
and check
the dependence of the numerical errors on the value of $\rho_{\rm out}$.

On the brane $\chi=\pi/2$, the boundary conditions are given by 
Israel's junction condition 
%===========<Equation>============%
%
\begin{equation}
K_{\mu\nu}=-\frac{1}{\ell}\gamma_{\mu\nu},
\label{junctioncondition}
\end{equation}
%
%=================================%
where $K_{\mu\nu}$ and $\gamma_{\mu\nu}$ are
the extrinsic curvature and the induced metric on the brane, respectively.
This condition implies that the brane has the constant tension
$\sigma=3/(4\pi G_5\ell)$. 
The condition \eqref{junctioncondition} is expressed as
%===========<Equation>============%
%
\begin{equation}
\frac{T_{,\chi}}{T}=R_{,\chi}=C_{,\chi}=\frac{\rho}{\ell}(e^{R}-1)
\label{braneboundary}
\end{equation}
%
%=================================%
in terms of the metric functions, and the constraint equation~\eqref{EE4}
is manifestly satisfied on the brane by these boundary conditions.

At the symmetry axis $\chi=0$, 
we impose the regularity conditions
%===========<Equation>============%
%
\begin{equation}
T_{,\chi}=R_{,\chi}=C_{,\chi}=0,
\label{BC:symmetry2}
\end{equation}
%
%=================================%
and
%===========<Equation>============%
%
\begin{equation}
R=C.
\label{BC:symmetry1}
\end{equation}
%
%=================================%
The second condition \eqref{BC:symmetry1}
is implied by the regularity of  
the second terms in Eqs.~\eqref{EE2} and \eqref{EE3}. 
Under the conditions \eqref{BC:symmetry1} 
and \eqref{BC:symmetry2}, 
the first constraint equation \eqref{EE4}
is trivially satisfied at $\chi=0$. Substituting
Eqs.~\eqref{BC:symmetry1} and \eqref{BC:symmetry2} 
into Eqs.~\eqref{EE1}--\eqref{EE3},
we obtain the formulas for $T_{,\chi\chi}$, $R_{,\chi\chi}$, 
and $C_{,\chi\chi}$. The second constraint equation~\eqref{EE5} 
evaluated at $\chi=0$
is consistent with these formulas.
In our numerical calculation, all these conditions are imposed
as explained in Sec.~\ref{Sec:IIIA}.

On the horizon $\rho=\rho_h$, the timelike killing vector should become null,
and thus
%===========<Equation>============%
%
\begin{equation}
T=0.
\label{BC:horizon1}
\end{equation}
%
%=================================%
Then, the regularity of  
Eqs~\eqref{EE2}--\eqref{EE5} implies
%===========<Equation>============%
%
\begin{equation}
C_{,\rho}=-\frac{\ell}{z\rho};
\label{BC:horizon2}
\end{equation}
%
%=================================%
%===========<Equation>============%
%
\begin{equation}
T_{,\rho\chi}=R_{,\chi}T_{,\rho};
\label{BC:horizon3}
\end{equation}
%
%=================================%
and
%===========<Equation>============%
%
\begin{equation}
2R_{,\rho}+C_{,\rho}=-\frac{3\ell}{z\rho}.
\label{BC:horizon4}
\end{equation}
%
%=================================%
Here, the condition \eqref{BC:horizon3}
is equivalent to the zeroth law of the black hole
thermodynamics, i.e. the constancy of the surface
gravity
%===========<Equation>============%
%
\begin{equation}
\kappa=e^{-R}T_{,\rho}
\end{equation}
%
%=================================%
on the horizon,
and the condition~\eqref{BC:horizon4} is equivalent 
to the zero expansion of the null geodesic
congruence on the horizon.
Here, it has to be pointed out that the four conditions
exist for the three variables.
If a regular solution exists, it has to satisfy all these
four boundary conditions on the horizon. But in the numerical
computation, only three conditions can completely fix the
solution to Eqs.~\eqref{EE1}--\eqref{EE3}. 
%In Ref.~\cite{KTN03}, Kudoh {\it et al.}
%basically used the conditions \eqref{BC:horizon1}, \eqref{BC:horizon2},
%and \eqref{BC:horizon4}, and used
%the integrated formula of \eqref{BC:horizon3}
%as a supplement condition to improve the numerical accuracy. 
The three conditions
\eqref{BC:horizon1}, \eqref{BC:horizon2}, and \eqref{BC:horizon4}
are used in our code,
since the use of \eqref{BC:horizon3}
leads to a numerical instability. 
Therefore, 
the constancy of the surface gravity $\kappa$ is not explicitly
imposed in our calculation, and whether
it is satisfied (within the expected
numerical error) has to be checked 
after generating a solution.
%This is one of the differences from the code by Kudoh {\it et al.}
%(although this is a small point).

%
%======================================%
%<<<<<<<<<<<< SECTION II  >>>>>>>>>>>>>>%
%======================================%
%
\section{Numerical method}
\label{Sec:III}

In this section, we explain how to solve the problem
numerically. Since our method is significantly improved
compared to the previous study by Kudoh {\it et al.} \cite{KTN03},
we focus attention to the differences between the two codes,
as summarized in Table~\ref{table1}.

%===========<Table1>============%
%
\TABLE[t]{
\begin{tabular}{c|cc}
\hline\hline
  &~~~~~~~~~~~Our code~~~~~~~~~~~&~~~~~~The code of Ref. \cite{KTN03}~~~~~~ \\
  \hline
radial coordinate & $x:=\log\rho$ & $\rho$ (geometric progression) \\
angular coordinate & $\chi$ & $\xi:=\chi^2$ \\
scheme & almost 4th-order & 2nd-order \\
grid size & $\Delta x=0.025$--$0.05$ & $\Delta\rho_1=0.049$ \\
grid error & $\sim 10^{-7}$--$10^{-5}$,  & $\sim 10^{-2}$ \\
outer boundary & $4\le x_{\rm out}\lesssim 7$ & $\rho_{\rm out}=85$\\
variables & $T,X,Y$ & $T,R,C$\\
\hline\hline
\end{tabular}
\caption{Comparison between our code and the code by 
Kudoh {\it et al.}~\cite{KTN03}.}
\label{table1}
}
%
%=================================%

%
%======================================%
%<<<<<<<<<< subsection 3.1 >>>>>>>>>>>>%
%======================================%
%
\subsection{Coordinates, variables, and scheme}
\label{Sec:IIIA}

In the code of Ref.~\cite{KTN03},
the nonuniform angular coordinate $\xi=\chi^2$ was adopted, 
because the second terms proportional to $1/\sin^2\chi$
in Eqs.~\eqref{EE2} and \eqref{EE3}
cause a severe numerical instability and the introduction of
$\xi$ makes the problem more tractable. But 
in this coordinate, 
the grid size measured in $\chi$ is not so small
in the neighborhood of the symmetry axis. For the grid number
100 in \cite{KTN03}, the grid next to the symmetry axis is located at 
$\chi=\pi/20\simeq 0.16$, and
this leads to a large numerical error.
For this reason, we adopt the uniform angular coordinate $\chi$ in our code. 
In this case, some treatment is
required in order to realize the numerical stability,
as explained in the next subsection.

The choice of the radial coordinate
is also different. The authors of \cite{KTN03} adopted
the method of the geometric progression, 
in which the grid size becomes larger
as the value of $\rho$ is increased.
They put
the outer boundary at $\rho_{\rm out}=85$ and used the grid number 1000,
where the ratio of the last grid size to the first grid 
size was set as $\Delta\rho_{\rm max}/\Delta\rho_1\simeq 2.71$
with $\Delta\rho_1=0.049$. 
For this choice, the expected grid error is 
$\sim (\Delta\rho_1)^2\simeq 0.24\%$, and
it is enhanced to $\sim 1\%$
in the neighborhood of the symmetry axis.
This is not so good as a numerical calculation. 
In our code, we use the
nonuniform radial coordinate $x$ defined by
%===========<Equation>============%
%
\begin{equation}
x:=\log(\rho/\rho_h).
\end{equation}
%
%=================================%
This means that the grid size measured in $\rho$ is
proportional to $\rho$.
However, this coordinate choice does not cause a significant
numerical error, because the error
becomes large only if the functions change rapidly,
while the
functions $T, R,$ and $C$ asymptote to the constant values
at the distant region. To be more precise, 
the functions that decay as $\sim 1/\rho$
behaves as $\sim \exp(-x)$ in the coordinate $x$, and
such an exponential decay is numerically tractable.
For this reason, we obtain the 
sufficient numerical accuracy with
relatively small grid numbers.
In our calculation, the outer boundary is located at $x_{\rm out}=10$ 
(i.e. $\rho_{\rm out}\simeq 22026.5$) for 
the case of a zero tension brane (Sec.~\ref{Sec:IV}).
In the cases of a nonzero tension brane,
we change the value of $x_{\rm out}$ around $5$ 
(i.e. $\rho_{\rm out}\simeq 148.4$)
and observe the dependence of the error on $x_{\rm out}$
(Secs.~\ref{Sec:VA} and \ref{Sec:VB}).

Let us turn to the choice of the variables.
In Ref.~\cite{KTN03}, the variables $T$, $R$, and $C$ 
were solved in the numerical calculation.
In our code, we introduce
%===========<Equation>============%
%
\begin{equation}
X:=R+C; \qquad Y:=R-C,
\end{equation}
%
%=================================%
and solve $T$, $X$, and $Y$. There are two reasons for this choice.
The first reason is related to the numerical stability. 
From Eqs.~\eqref{EE2} and ~\eqref{EE3}, we derive the equations 
of the form 
$\nabla^2 X=\cdots$ and $\nabla^2 Y=\cdots$. In these equations,
the term proportional to $1/\sin^2\chi$ is included only 
in the latter equation, and thus
it becomes easier to handle the instability caused
by this term.
The second reason is that
the nonsystematic error in the case 
of a nonzero tension brane appears as an unnatural jump
in the value of $Y$ in the neighborhood of the axis.

To summarize, the variables $T$, $X$, and $Y$
are solved using the coordinates
$x$ and $\chi$ in our code. 
We adopt the finite difference method, where
the grids are located at $x=I\Delta x$ ($I=0,...,I_{\rm max}$)
and $\chi=J\Delta \chi$ ($J=0,...,J_{\rm max}$).
Here, $\Delta x:=x_{\rm out}/I_{\rm max}$
and $\Delta \chi:=(\pi/2)/J_{\rm max}$,
and we keep the ratio $\Delta \chi/\Delta x=\pi/2$.
We adopt the 4th-order accuracy scheme except
at the grids in the neighborhood of the three boundaries 
$\chi=\pi/2$, $x=0$, and $x=x_{\rm out}$, where
the 3rd-order accuracy scheme is adopted.
Therefore, 
our numerical code has the almost 
4th-order accuracy, 
and therefore we can obtain more accurate results
with smaller grid numbers
compared to the case that the 2nd-order accuracy 
scheme is adopted. As proved in next section,
the error by the finiteness of the grid size
(say, the {\it ``grid error''}) is $\lesssim 10^{-5}$
for $\Delta x=0.05$ and $\sim 10^{-7}$ for $\Delta x=0.025$.
Compared to the grid error $\sim 1\%$ in Ref.~\cite{KTN03},
the numerical accuracy is greatly improved.

We briefly mention how to impose the boundary conditions at the symmetry
axis.
There, we have the regularity conditions $Y=0$, 
$T_{,\chi}=X_{,\chi}=Y_{,\chi}=0$, and formulas for 
$T_{,\chi\chi}$, $X_{,\chi\chi}$, and $Y_{,\chi\chi}$.
It is possible to write down the finite difference
equations for $T$, $X$, and $Y$ at $J=0$ (i.e. $\chi=0$)
that include the conditions for the first and second order
derivatives.
We adopt those finite difference equations for $T$ and $X$ 
in order to determine the 
values of $T$ and $X$ at $J=0$.
As for $Y$, we impose $Y=0$ at $J=0$, and
use the remaining finite difference equation for $Y$ 
to determine the value of $Y$ at $J=1$ (i.e. $\chi=\Delta\chi$).  
This is better than using the finite difference equation
at $J=1$ that has the term proportional to $1/\sin^2\chi$.

%
%======================================%
%<<<<<<<<<< subsection 3.2 >>>>>>>>>>>>%
%======================================%
%
\subsection{The numerical relaxation}
\label{Sec:IIIB}

In order to solve the finite difference equations,
we adopt the relaxation method,
in which one prepares an initial surface and
makes it converge to the solution iteratively. 
Our code is 
based on the successive-over-relaxation (SOR) method.
Namely, we calculate the difference from the finite difference
equations $\Delta T_{(I,J)}$,  $\Delta X_{(I,J)}$,  
and $\Delta Y_{(I,J)}$, and determine the next surface
using the formulas
%===========<Equation>============%
%
\begin{eqnarray}
T^{\rm (next)}_{(I,J)}&=&T_{(I,J)}+w_T\Delta T_{(I,J)};
\nonumber\\
X^{\rm (next)}_{(I,J)}&=&X_{(I,J)}+w_X\Delta X_{(I,J)};
\\
Y^{\rm (next)}_{(I,J)}&=&Y_{(I,J)}+w_Y\Delta Y_{(I,J)}.
\nonumber
\end{eqnarray}
%
%=================================%
In the usual SOR method, $w_T$, $w_X$, and $w_Y$ are called 
the acceleration parameters
and are chosen to be a value between $1$ and $2$.
In our case, in order to realize the stability,
we choose
%===========<Equation>============%
%
\begin{equation}
w_T=w_X=0.8; \qquad w_Y=0.08\times\cos^{\alpha}[(x-x_{\rm out})\pi/2]\sin\chi,
\label{SOR-parameters}
\end{equation} 
%
%=================================%
where $\alpha$ is chosen appropriately depending
on the situation. 
The reason for the complicated functional form of $w_Y$
is as follows. The numerical instability tends to happen
at the distant region or near the symmetry axis.
In those regions, 
the value of $w_Y$ is very small and 
this makes the convergence of the surface
very slow. In this way, 
we can avoid the instability 
caused by the term proportional to $1/\sin^{2}\chi$.

In order to evaluate the degree of the convergence,
we introduce the parameters
%===========<Equation>============%
%
\begin{equation}
\epsilon_T=
\frac{\sum \left|\Delta T_{(I,J)}\right|}
{\sum\left|T_{(I,J)}\right|};
\qquad 
\epsilon_X=
\frac{\sum\left|\Delta X_{(I,J)}\right|}
{\sum\left|X_{(I,J)}\right|};
\qquad 
\epsilon_Y=
\frac{\sum\left|\Delta Y_{(I,J)}\right|}
{I_{\rm max}J_{\rm max}}.
\end{equation}
In the formula of $\epsilon_Y$,
the normalization factor was not adopted as $\sum{Y_{(I,J)}}$
because the analytic solution is $Y=0$ in the
case of a zero tension brane.
Typically, the value of $\epsilon_Y$ is the largest among the
three convergence parameters. 
We truncated the relaxation process
when 
%===========<Equation>============%
%
\begin{equation}
\max (\epsilon_T,\epsilon_X, \epsilon_Y)<\epsilon_0
\label{convergence-criteria}
\end{equation}
is achieved. 
In the cases of a nonzero tension brane, 
the value of $\epsilon_0$ is taken as
$\epsilon_0=10^{-10}$. 
The error from this truncation (say, the {\it ``relaxation error''}) 
is $\sim 10^{-6}$.
In the case of a zero tension brane in the next section,
we adopt $\epsilon_0=10^{-12}$ where 
the relaxation error is $\sim 10^{-8}$,
because we have to make the relaxation error 
sufficiently smaller than the grid error
in order to prove the appropriate convergence 
of the numerical solutions.

%
%======================================%
%<<<<<<<<<<<< SECTION IV  >>>>>>>>>>>>>>%
%======================================%
%
\section{Results for a zero tension brane}
\label{Sec:IV}

In this section, we show the results for the case of a zero tension brane,
i.e. the spherically symmetric case.
In this case, $(1/L)=0$ 
and the analytic solution~\eqref{spherical-case} (i.e.,
$T_0$, $R_0$, and $C_0$)
exists. A comparison of the numerical data with the analytic solution
gives a useful check for the correctness of our code.
Furthermore, we can learn the features that the numerical
solutions have to satisfy when the solutions are actually present.

In these calculations, we set $\rho/\ell=0$
in our code. Except this,
we do not explicitly impose the spherical symmetry.
As a result, the numerical solution slightly depends on $\chi$
because of the numerical errors.
It is worth pointing out that 
the sufficient convergence
was realized with the initial surface
that depends on $\chi$, in spite of the presence of
the term proportional
to $1/\sin^2\chi$.
This proves 
the sufficient stability of our code.
We calculated $T$, $X$ and $Y$ for various locations 
of the outer boundary $x_{\rm out}$ 
and grid sizes $\Delta x$.
We changed the value of $x_{\rm out}$
from $3$ to $10$ and obtained the natural result
that the solution converges as $x_{\rm out}$ is increased.

%===========<Figure1>============%
%
\FIGURE[t]{\centerline
{
\epsfig{file=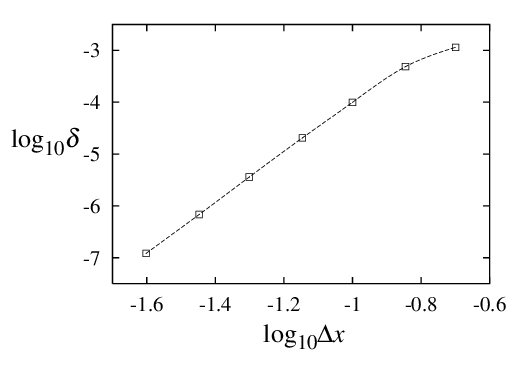,width=0.50\textwidth}
}
\caption{The relation between the grid size $\Delta x$ 
and the error $\delta$ in the case of a zero tension brane 
in the log scales.
The numerical data shows the almost 4th-order convergence.
}
\label{converge4}
}
%
%=================================%

It is important to confirm 
the appropriate convergence of the numerical data. 
For this purpose, we change the grid size $\Delta x$
from $0.025$ to $0.2$
for a fixed  $x_{\rm out}=10$.
For this choice of $x_{\rm out}$,
the boundary error is smaller
than the grid error.
Figure~\ref{converge4}
shows the relation between $\Delta x$ and the
error defined by
%===========<Equation>============%
%
\begin{equation}
\delta:=\frac{\sum_{I,J}\left(|T-T_0|+|X-X_0|+|Y-Y_0|\right)}{
\sum_{I,J}\left(|T_0|+|X_0|+|Y_0|\right)}.
\end{equation}
%
%=================================%
From this figure, 
the almost 4th-order convergence is confirmed.
In fact, the slope of the curve in Fig.~\ref{converge4} is
$\sim 4.8$, a bit larger than $4$.
This is interpreted as follows.
Since we partly use the 3rd-order accuracy scheme, 
the error is expected to be between
$O(\Delta x^3)$ and $O(\Delta x^4)$.
As $\Delta x$ is decreased,
the ratio of the number of the 3rd-order grids
to the number of the 4th-order grids 
becomes smaller and this change in the ratio
makes the error from the 3rd-order
grids less effective. As a result,
the slope of the curve becomes larger than 4 in Fig.~\ref{converge4}.

%===========<Figure2>============%
%
\FIGURE[t]{
\centerline
{
\epsfig{file=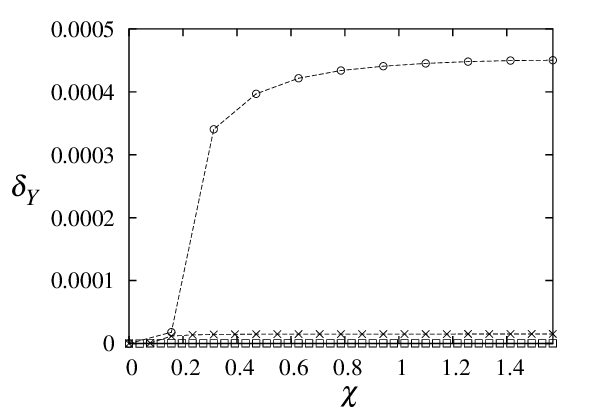,width=0.50\textwidth}
}
\caption{The angular dependence of the error $\delta_Y(\chi)$
in $Y$ for $\Delta x=0.1$ ($\circ$), 
$0.05$ ($\times$), 
and $0.025$ ($\square$).
The error shows the almost 4th-order convergence and
is almost invisible for the case $\Delta x=0.025$.
The value of $\delta_Y(\chi)$ tends to decrease as the value of $\chi$
is decreased.}
\label{errorang-spherical}
}
%
%=================================%

Let us discuss the angular dependence of the numerical error
in $Y$, since it will be important in the
case of a nonzero tension brane.
Since $R_0=C_0$ and thus $Y_0=0$, the numerical value of
$Y$ itself represents the error.
For this reason, we define
%===========<Equation>============%
%
\begin{equation}
\delta_Y(\chi):=\max[Y]_\chi-\min[Y]_\chi
\end{equation}
%
%=================================%
as the error in $Y$ characteristic to 
the angular direction $\chi$. Here,
$\max[Y]_\chi$ and $\min[Y]_\chi$
mean the maximum and minimum values of $Y$ for a given
coordinate value $\chi$.
Figure~\ref{errorang-spherical}
shows the behavior of $\delta_Y(\chi)$
for the grid size $\Delta x=0.1$, $0.05$, and $0.025$.
The almost 4th-order convergence is again confirmed.
Moreover, it is seen that $\delta_Y(\chi)$
is suppressed in the neighborhood of the symmetry axis.
It is important to point out that this tendency is 
in contrast to the case of the 2nd-order code.
In that case, the value of
$Y$ tends to increase as the value of $\chi$ is decreased,
and suddenly jumps to $Y\simeq 0$ at $\chi=\Delta\chi$
because of the imposed boundary condition.
Such an unnatural behavior
does not appear in the 4th-order code, and 
it is one of the merits in using the 4th-order accuracy scheme.

To summarize, we have proved that our code
has the sufficient numerical stability and 
successfully reproduces the solution \eqref{spherical-case}
in the zero tension case.
Our result shows that: (i) The numerical solution converges
as the value of $x_{\rm out}$ is increased;
(ii) The numerical solution shows the almost 4th-order convergence;
and (iii) The numerical error in $Y$ is suppressed in the
neighborhood of the symmetry axis $\chi=0$. 
These three features are expected to be held
also for the cases where the brane tension
is nonzero as long as $(1/L)$ is sufficiently small
and regular solutions exist.

Here, we introduce 
{\it ``systematic error''} and {\it ``nonsystematic error,''}
which play important roles in the next section.
The systematic error means the error
that originates from the adopted approximation
in the numerical method. The systematic errors were
already introduced in this paper: the ``boundary error''
from the finiteness of the location of the outer boundary $x_{\rm out}$,
the ``grid error'' from the finiteness of the grid sizes,
and the ``relaxation error'' from the truncation of
the numerical relaxation. In our method, there is no other source
of the numerical error. We can avoid the relaxation error
by making the criterion of the truncation strict. The boundary 
error has to decrease as $x_{\rm out}$ is increased,
and the grid error has to decrease consistently with
the adopted scheme as $\Delta x$ is decreased. 
In other words, when the numerical convergence is sufficient,
the obtained solution has to satisfy the features (i) and (ii)
above. 
If we detect an error that does not satisfy the features (i) and (ii),
we call it the nonsystematic error.
If such a nonsystematic error is detected, we have to suspect
that something is wrong.

%
%======================================%
%<<<<<<<<<<<< SECTION V  >>>>>>>>>>>>>>%
%======================================%
%
\section{Results for a nonzero tension brane}
\label{Sec:V}

In this section, we show the numerical results
for the case where the brane tension is nonzero.
In Secs.~\ref{Sec:VA} and \ref{Sec:VB}, the errors in $Y$ and the surface
gravity $\kappa$ are discussed in detail, respectively, 
and the existence of the nonsystematic error is shown. 
In Sec.~\ref{Sec:VC}, we compare our results with the perturbative study 
\cite{KSSW04} ignoring the numerical errors.
In Sec.~\ref{Sec:VD}, the interpretation of our numerical results 
is discussed.

%
%======================================%
%<<<<<<<<<< subsection 5.1 >>>>>>>>>>>>%
%======================================%
%
\subsection{The error in the variable $Y$}
\label{Sec:VA}

There is a limitation in the values of 
$(1/L)$ and $x_{\rm out}$ for which 
numerical solutions can be obtained.
For a fixed $(1/L)$, the numerical relaxation
is stable only for 
$x_{\rm out}\le x_{\rm out}^{\rm (crit)}(1/L)$
for some critical value $x_{\rm out}^{\rm (crit)}(1/L)$.
The value of $x_{\rm out}^{\rm (crit)}(1/L)$
becomes smaller as $(1/L)$ is increased
and depends also on the resolution. 
When the numerical instability does not occur,
we always realized the criterion \eqref{convergence-criteria} 
for the truncation of the relaxation.

%===========<Figure3>============%
%
\FIGURE[t]{
\centerline
{
\epsfig{file=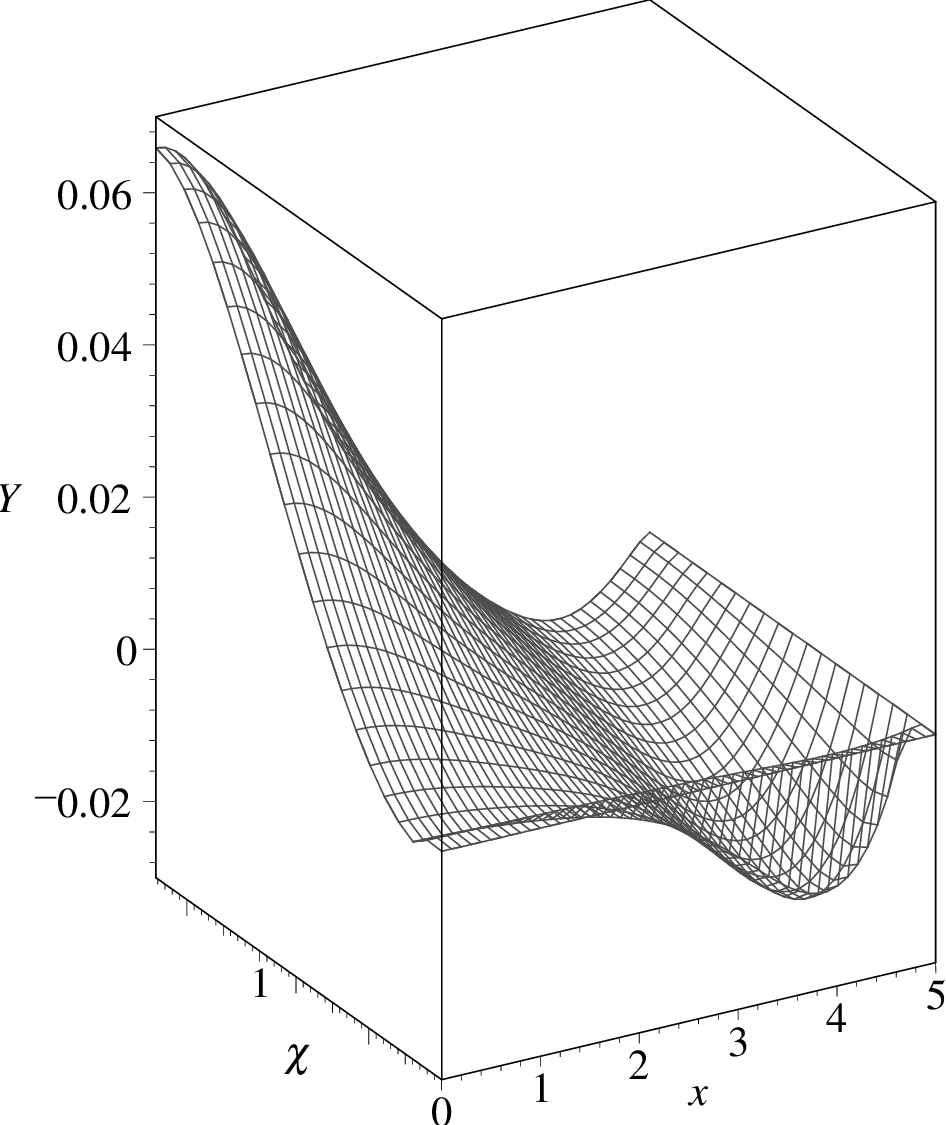,width=0.40\textwidth}
}
\caption{The numerical solution of $Y$ for $(1/L)=0.06$
as a function of $x$ and $\chi$. Here, $x_{\rm out}=5$ 
and $\Delta x=0.05$ are adopted.
The value of $Y$ does not decay properly at the distant region.
Besides, there is an unnatural jump in the value of $Y$ between the
grids $J=1$ and $2$ (i.e. $\chi=\Delta\chi$ and $2\Delta\chi$).
}
\label{threeD-L16.6-Y}
}
%
%=================================%
%===========<Figure3>============%
%
%\EPSFIGURE
%{fig3.eps,width=0.40\textwidth}
%{The numerical solution of $Y$ for $(1/L)=0.06$
%as a function of $x$ and $\chi$. Here, $x_{\rm out}=5$ 
%and $\Delta x=0.05$ are adopted.
%The value of $Y$ does not decay properly at the distant region.
%Besides, there is an unnatural jump in the value of $Y$ between the
%grids $J=1$ and $2$ (i.e. $\chi=\Delta\chi$ and $2\Delta\chi$).
%\label{threeD-L16.6-Y}
%}
%
%=================================%

When both $(1/L)$ and $x_{\rm out}$ are not so large
(e.g. $(1/L)=0.02$ and $x_{\rm out}=5.0$),
the numerical solutions for $T$, $X$, and $Y$ 
(apparently) look natural,
and their gross features coincide
with those of the solution in \cite{KTN03}. 
For example, the figures in \cite{KTN03} indicate that
$Y\ge 0$ and $Y$ takes the maximum value where the
brane and the horizon cross each other, and the same feature
holds in our numerical data. However, as $(1/L)$
is increased for a fixed $x_{\rm out}$, a strange behavior becomes 
relevant.
Figure~\ref{threeD-L16.6-Y} shows our numerical solution of 
$Y(x,\chi)$ for $(1/L)=0.06$, $x_{\rm out}=5.0$, 
and $\Delta x=0.05$. 
The value of $Y$ does not decay properly at the distant region.
Moreover, there is an unnatural jump in the values of $Y$
at the grids $J=1$ and $2$ (i.e. $\chi=\Delta\chi$ and $2\Delta\chi$).
This is in contrast to the case of a zero tension brane
in Sec.~\ref{Sec:IV},
where the error in $Y$ is suppressed in the neighborhood
of the axis.  
As the characteristic error in $Y$, 
we define the following quantity:
%===========<Equation>============%
%
\begin{equation}
\delta_Y:=\max\left|Y(x,2\Delta\chi)-Y_{\rm ext}(x)\right|.
\end{equation}
%
%=================================%
Here, $Y_{\rm ext}(x)$ means the extrapolated value of $Y$
at $\chi=2\Delta\chi$ using the values of $Y(x,0)$ and $Y(x,\Delta\chi)$.
Let us observe how the value of $\delta_Y$ depends on $x_{\rm out}$
and $(1/L)$.

%===========<Figure4>============%
%
\FIGURE[t]{
\centerline
{
\epsfig{file=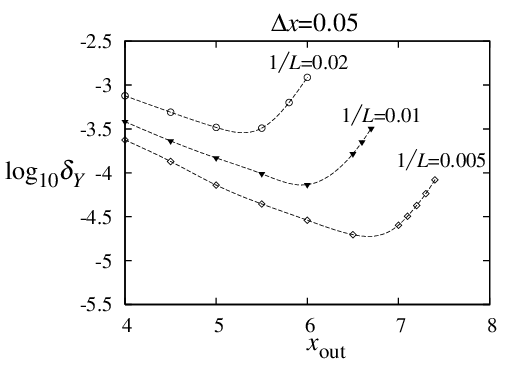,width=0.49\textwidth}
\epsfig{file=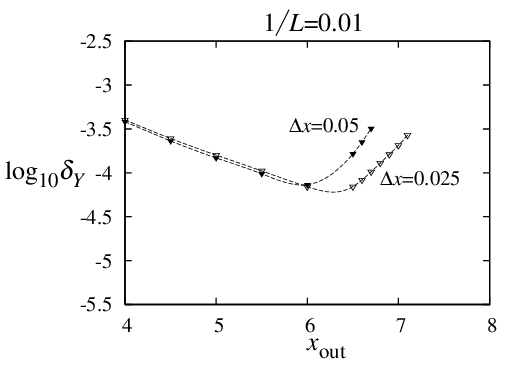,width=0.49\textwidth}
}
\caption{Dependence of the error $\delta_Y$ in $Y$ on
the location of the outer boundary $x_{\rm out}$.
The left plot shows the cases of $(1/L)=0.005$ ($\circ$), 
$0.01$ ($\blacktriangledown$),
and $0.02$ ($\diamond$) for a fixed resolution $\Delta x=0.05$. 
The right plot shows the 
results by the two different resolutions, $\Delta x=0.05$ ($\blacktriangledown$)
and $0.025$ ($\triangledown$),
in the case $(1/L)=0.01$. 
In the range $x_{\rm out}\lesssim \log(4L)$,
the error value $\delta_Y$
does not depend on the resolution and 
decreases 
with the increase in $x_{\rm out}$.
In the range $x_{\rm out}\gtrsim \log(4L)$,
the error value $\delta_Y$
depends on the resolution and
becomes larger with the increase in $x_{\rm out}$ in both resolutions.
The error is interpreted as the boundary error and
the nonsystematic error for $x_{\rm out}\lesssim \log(4L)$
and $x_{\rm out}\gtrsim \log(4L)$, respectively.
See text for details.
}
\label{xout-log10deltaY}
}
%
%=================================%

Figure~\ref{xout-log10deltaY} shows the
value of $\delta_Y$ as a function of $x_{\rm out}$. 
In the left plot, the cases $(1/L)=0.005$, $0.01$, and $0.02$
are shown for a fixed $\Delta x=0.05$.
As $x_{\rm out}$ is increased,
$\delta_Y$ decreases at first, but later
increases. The change from decrease to increase
happens at $x_{\rm out}\simeq\log(4L)$
(i.e. $\rho_{\rm out}\simeq 4\ell$). 
This behavior is opposite to our expectation,
since the error has to continue to decrease as $x_{\rm out}$ is increased
until it hits one of the relaxation and grid errors which are both
$\lesssim 10^{-5}$. 
Therefore, although the error $\delta_Y$ for $x_{\rm out}\lesssim \log(4L)$
can be interpreted as the boundary error,
we have detected the nonsystematic error
in the range $x_{\rm out}\gtrsim \log(4L)$.
The right plot shows the
comparison between the two resolutions, $\Delta x=0.05$
and $0.025$, for a fixed $(1/L)=0.01$. 
For $x_{\rm out}\lesssim \log(4L)$, the two error values
almost coincide and they are determined only by $x_{\rm out}$. 
Therefore, we can confirm that the error in this region
is the boundary error.
On the other hand, for $x_{\rm out}\lesssim \log(4L)$, the error $\delta_Y$
depends on the resolution.
However, they are not the grid errors, since the grid errors
are $\lesssim 10^{-5}$ and $\sim 10^{-7}$ for
$\Delta x=0.05$ and $0.025$, respectively,
as proved in Sec.~\ref{Sec:IV}. 
The relaxation error is also $\sim 10^{-6}$.
The important feature is that in both resolutions,
the errors become larger as $x_{\rm out}$ is increased.
Therefore, we again confirm that the error 
in the range $x_{\rm out}\gtrsim \log(4L)$ is the nonsystematic error.

%===========<Figure5>============%
%
\FIGURE[t]{
\centerline
{
\epsfig{file=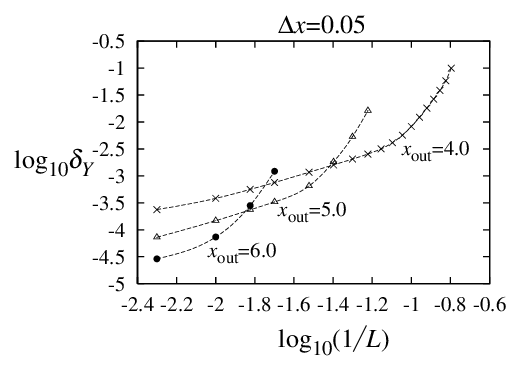,width=0.49\textwidth}
\epsfig{file=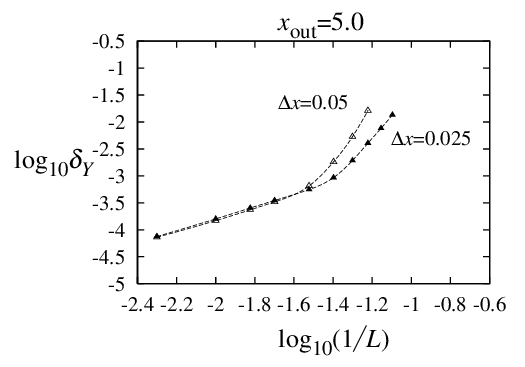,width=0.49\textwidth}
}
\caption{Dependence of the error $\delta_Y$ in $Y$ on
the values of $(1/L)$.
The left plot shows the cases of $x_{\rm out}=4.0$ ($\times$), 
$5.0$ ($\vartriangle$), and $6.0$ ($\bullet$) 
for a fixed resolution $\Delta x=0.05$. 
The right plot shows the results by the 
two different resolutions, $\Delta x=0.05$ ($\vartriangle$) 
and $0.025$ ($\blacktriangle$),
in the case $x_{\rm out}=5.0$. 
The boundary error $\delta_Y$ 
in the range $(1/L)\lesssim 4\exp(-x_{\rm out})$ 
scarcely depends on the resolution
and grows linearly with respect to $(1/L)$. 
The nonsystematic error 
in the range $(1/L)\gtrsim 4\exp(-x_{\rm out})$ depends on the 
resolution and grows nonlinearly with respect to $(1/L)$ 
in both resolutions.}
\label{log10IL-log10deltaY}
}
%
%=================================%

Figure~\ref{log10IL-log10deltaY} shows the value of  $\delta_Y$ 
as a function of $(1/L)$.
The left plot shows the cases of $x_{\rm out}=4.0$, 
$5.0$, and $6.0$ for a fixed $\Delta x=0.05$. 
%Again, we can confirm that the boundary error
%decreases as $x_{\rm out}$ is increased.
We see 
the boundary error in the range $(1/L)\lesssim 4\exp(-x_{\rm out})$,
and it grows linearly
with respect to $(1/L)$ for a fixed $x_{\rm out}$. 
This is natural 
because the value of $Y$ itself grows linearly with $(1/L)$.
On the other hand, 
we see the nonsystematic error
in the range $(1/L)\gtrsim 4\exp(-x_{\rm out})$.
In contrast to the boundary error,
the nonsystematic error grows nonlinearly with respect to $(1/L)$
for a fixed $x_{\rm out}$,
and the growth rate is roughly given by $\sim (1/L)^4$.
The right plot shows the comparison between 
the two different resolutions, $\Delta x=0.05$ and $0.025$,
for a fixed $x_{\rm out}=5.0$. 
Again, we confirm that the boundary errors are almost same 
for both resolutions, while the nonsystematic error 
depends on the resolution. 
Although the nonsystematic error for $\Delta x=0.025$ 
is smaller than that for $\Delta x=0.05$ if compared
with the same $(1/L)$, 
the nonlinear growth $\sim (1/L)^4$ is observed in both resolutions.

%
%======================================%
%<<<<<<<<<< subsection 5.2 >>>>>>>>>>>>%
%======================================%
%
\subsection{The error in the surface gravity $\kappa$}
\label{Sec:VB}

%===========<Figure6>============%
%
\FIGURE[t]{
\centerline
{
\epsfig{file=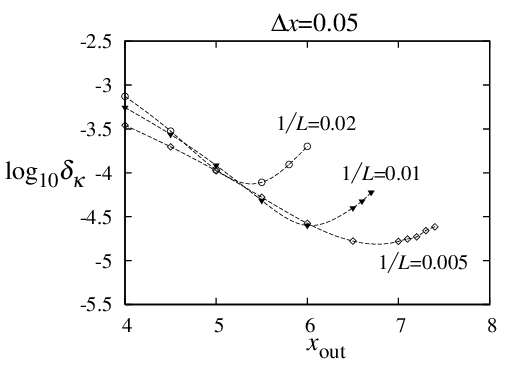,width=0.49\textwidth}
\epsfig{file=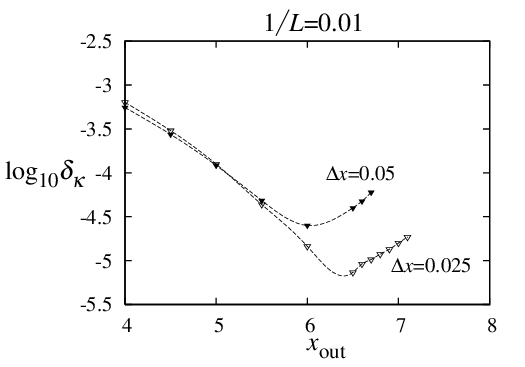,width=0.49\textwidth}
}
\caption{The same as Fig.~\ref{xout-log10deltaY} but for
the error $\delta_\kappa$ in the surface gravity. The
boundary error and the nonsystematic error can be
seen for $x_{\rm out}\lesssim \log(4L)$ and $x_{\rm out}\gtrsim \log(4L)$,
respectively, also in the error $\delta_{\kappa}$.}
\label{xout-log10deltakappa}
}
%
%=================================%

%===========<Figure7>============%
%
\FIGURE[t]{
\centerline
{
\epsfig{file=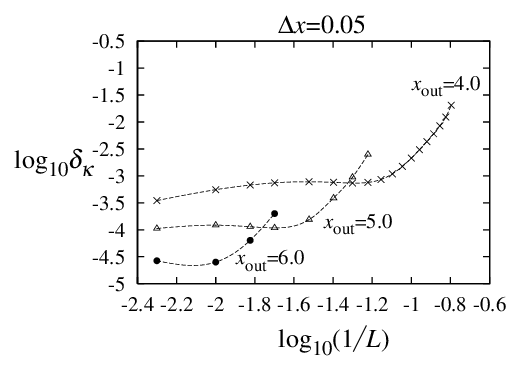,width=0.49\textwidth}
\epsfig{file=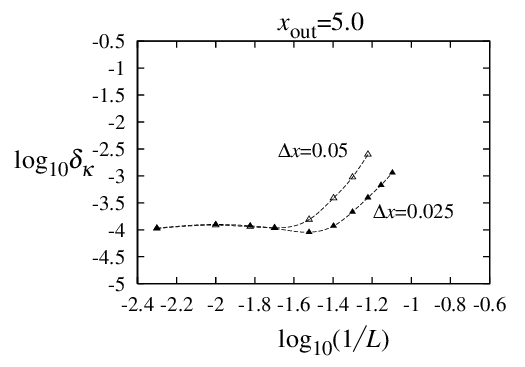,width=0.49\textwidth}
}
\caption{The same as Fig.~\ref{log10IL-log10deltaY}
but for the error $\delta_{\kappa}$ in the surface gravity.
The nonlinear growth of the nonsystematic error
with respect to $(1/L)$ is seen also for $\delta_{\kappa}$.}
\label{log10IL-log10deltakappa}
}
%
%=================================%

As another example of the numerical error,
we look at the error in the surface gravity $\kappa$. 
As mentioned in Sec.~\ref{Sec:IIB}, there are the four boundary conditions
on the horizon, and the three of them are chosen in our numerical
calculation. Since we omitted the boundary condition
that corresponds to the constancy of $\kappa$ on the horizon,
we have to check if it is satisfied within the
amount of the systematic errors after generating the solutions.
In order to evaluate the error in $\kappa$, we define
%===========<Equation>============%
%
\begin{equation}
\delta_{\kappa}:=\max[\kappa]-\min[\kappa].
\end{equation}
%
%=================================%
Figure~\ref{xout-log10deltakappa}
shows the value of $\delta_{\kappa}$ as a function of $x_{\rm out}$.
The left plot shows the cases of $(1/L)=0.005$, 
$0.01$, and $0.02$ for a fixed $\Delta x=0.05$,
and the right plot shows the comparison between the two different
resolutions, $\Delta x=0.025$ and $0.05$, for a fixed $(1/L)=0.01$. 
Although $\delta_{\kappa}$
is smaller than $\delta_Y$,
they have common features. Namely, if $x_{\rm out}$ is increased,
the value of 
$\delta_{\kappa}$ decreases at first but begins to increase
at $x_{\rm out}\simeq \log(4L)$. 
Therefore, also for $\delta_{\kappa}$,
we find the boundary error and the nonsystematic error 
in the ranges $x_{\rm out}\lesssim \log(4L)$
and $x_{\rm out}\gtrsim \log(4L)$, respectively.
Although the nonsystematic errors
for the resolutions $\Delta x=0.05$ and $0.025$ 
have a little different values, they
both become larger as $x_{\rm out}$ is increased.

Figure~\ref{log10IL-log10deltakappa} shows the
value of $\delta_{\kappa}$ as a function of $(1/L)$.
The left plot shows the cases of $x_{\rm out}=4.0$, 
$5.0$, and $6.0$ for a fixed $\Delta x=0.05$,
and the right plot shows the comparison between
the two different resolutions, $\Delta x=0.05$ and $0.025$,
for a fixed $x_{\rm out}=5.0$. The boundary error for $\delta_{\kappa}$
is almost constant when $(1/L)$ is changed and this is 
different from the behavior of $\delta_Y\propto (1/L)$.
This is because the surface gravity $\kappa$ itself
is almost constant.
On the other hand, for a fixed $x_{\rm out}$,
the nonsystematic error for $\delta_{\kappa}$ 
grows nonlinearly with respect to $(1/L)$.
Although the nonsystematic error for $\Delta x=0.025$
is smaller than that for $\Delta x=0.05$,
we observe the nonlinear growth with respect to $(1/L)$
in both resolutions, and again the growth rate is roughly 
$\sim (1/L)^4$.

%
%======================================%
%<<<<<<<<<< subsection 5.3 >>>>>>>>>>>>%
%======================================%
%
\subsection{Comparison with the perturbative study}
\label{Sec:VC}

In the next subsection, we discuss the fact that the nonsystematic
error has a crucial meaning for the existence of the solution.
But in the perturbative level, the (at least $C^1$) solution
of a black hole on a brane was found for a small mass~\cite{KSSW03,KSSW04}.
In this subsection, let us examine whether our numerical solution
is consistent with the perturbative study 
if the numerical errors are ignored.
Since the perturbative study~\cite{KSSW03,KSSW04}
was done in a different gauge condition, 
the direct comparison between 
the metric functions is not possible. However, 
we can compare the two results 
by calculating nondimensional quantities on the horizon.

%===========<Figure8,9>============%
%
\DOUBLEFIGURE[t]
{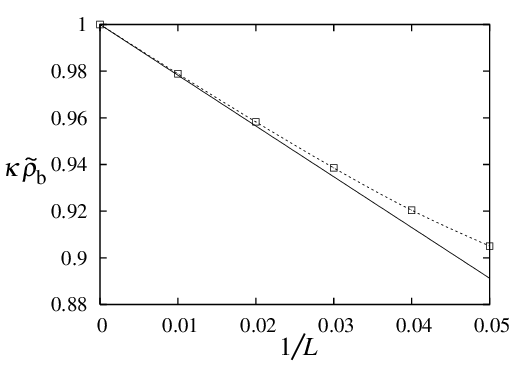,width=0.47\textwidth}
{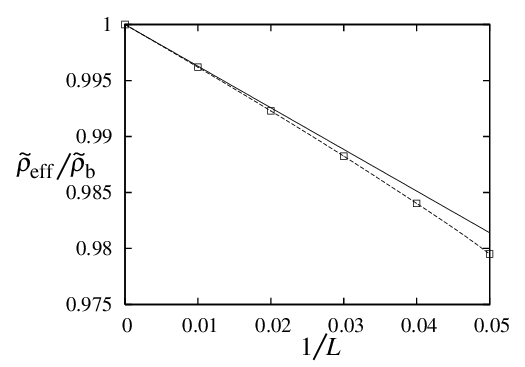,width=0.47\textwidth}
{The behavior of
$\kappa\tilde{\rho}_{\rm b}$ as a function of $(1/L)$.
The numerical data for $x_{\rm out}=5.0$ and $\Delta x=0.05$
is shown by squares ($\square$). The
prediction by the perturbative study \cite{KSSW04}
is shown by a solid line. The two results agree well for $(1/L)\ll 1$.
\label{sgxrad4}}
{The same as Fig.~8 %~\ref{sgxrad4} 
but for the quantity 
$\tilde{\rho}_{\rm eff}/\tilde{\rho}_{\rm b}$. 
Again, the two results agree well
for $(1/L)\ll 1$.
\label{effective_brane_radius}}
%
%=================================%

As the first example,
let us consider the product of the surface gravity $\kappa$
and the proper radius of the horizon on the brane
%===========<Equation>============%
% 
\begin{equation}
\tilde{\rho}_{\rm b}:=\rho_h\exp\left[C(\rho_h,\pi/2)\right]
\end{equation}
%
%=================================%
(i.e., the proper circumference divided by $2\pi$).
Here, $C$ is written as a function of the coordinates $(\rho,\chi)$.
Figure~\ref{sgxrad4} shows the relation between $(1/L)$
and $\kappa\tilde{\rho}_{\rm b}$
for fixed values of $\Delta x=0.05$ and $x_{\rm out}=5.0$.
The prediction by the perturbative study \cite{KSSW04}
is shown by a solid line (see Appendix A for a derivation). 
The two results agree well 
in the regime $(1/L)\ll 1$.

As another example, let us consider the quantity
related to the distortion of the horizon. 
For this purpose, we define the effective radius
of the horizon in the 5-dimensional sense as
%===========<Equation>============%
%
\begin{equation}
\tilde{\rho}_{\rm eff}:=\left({A_5}/{\Omega_3}\right)^{1/3},
\end{equation}
%
%=================================%
and calculate the ratio of the effective radius $\tilde{\rho}_{\rm eff}$
to the proper radius on the brane $\tilde{\rho}_{\rm b}$.
Here, $\Omega_3$ denotes the 3-dimensional area of
a unit sphere, $\Omega_3=2\pi^2$, and $A_5$
is the area of the horizon 
%===========<Equation>============%
%
\begin{equation}
A_5:=2\Omega_2\int_{0}^{\pi/2} 
\left(\frac{\ell}{z}\right)^3e^{R+2C}\sin^2\chi d\chi, 
\end{equation}
%
%=================================%
with $\Omega_2:=4\pi$.  
Figure~\ref{effective_brane_radius} shows the relation
between $(1/L)$ and $\tilde{\rho}_{\rm eff}/\tilde{\rho}_{\rm b}$
for fixed values of $\Delta x=0.05$ and $x_{\rm out}=5.0$.
The value of $\tilde{\rho}_{\rm eff}/\tilde{\rho}_{\rm b}$
decreases as $(1/L)$ is increased, and thus the horizon 
becomes flattened. The prediction by the perturbative study~\cite{KSSW04}
is shown by the solid line (see Appendix A for a derivation), 
and the numerical data 
agrees well with the perturbative study in the regime $(1/L)\ll 1$.

As found above, our numerical data is consistent
with the perturbative study~\cite{KSSW03,KSSW04}
if the numerical errors are ignored.
This result can be interpreted as the evidence for
the correctness of our code
as well as the correctness of the perturbative study.

%
%======================================%
%<<<<<<<<<< subsection 5.4 >>>>>>>>>>>>%
%======================================%
%
\subsection{Interpretation}
\label{Sec:VD}

Now, let us discuss the interpretation of our results.
In Secs.~\ref{Sec:VA} and \ref{Sec:VB}, 
we have found the nonsystematic
errors in our numerical data for both the
error $\delta_Y$ in $Y$ and the error $\delta_{\kappa}$
in the surface gravity. These nonsystematic
errors become relevant for $x_{\rm out}\gtrsim \log (4L)$,
and grow nonlinearly with respect to $(1/L)$ for a fixed $x_{\rm out}$.
On the other hand, as shown in Sec.~\ref{Sec:VC}, 
our results are consistent with the
perturbative study~\cite{KSSW03,KSSW04}
for a small black hole on a brane in the regime $(1/L)\ll 1$ 
if the numerical errors
are ignored. In order to interpret these strange results,
we suggest and discuss the following three possibilities:
(I) The numerical code is wrong; (II) The solution does not exist;
and (III) The solution exists but it is numerically unstable.

We understand that the possibility (I) is the case that
most readers might expect. However, we have successfully
reproduced the Schwarzschild solution for $(1/L)=0$
in Sec.~\ref{Sec:IV}. Moreover, our results
for $(1/L)\ll 1$ are
consistent with the perturbative study \cite{KSSW03,KSSW04}.
This guarantees that at least the linear terms in $\rho/\ell$ 
are correctly included in our code. 
Therefore, our task is to properly include the other
nonlinear terms, and this can
be done by careful checks. For this reason, we definitely exclude
the possibility (I).

We turn to the possibility (II) that
no solution exists. Here, we consider the case
that no regular solution exists but the solution
to the finite difference equations exist. 
Since such a case can happen, it is generally warned
that at least the appropriate convergence
has to be checked in numerical calculations. 
Let us recall Figs.~\ref{xout-log10deltaY},
\ref{log10IL-log10deltaY}, \ref{xout-log10deltakappa}, 
and \ref{log10IL-log10deltakappa}. In the right plots of those figures,
we compared the errors of the results by the two
different resolutions, $\Delta x=0.05$ and $0.025$,
and found that the numerical data does not show the
appropriate convergence. If we assume that the solution
to the finite difference equations is unique (see possibility (III)
for the meaning of this assumption), this indicates that a
regular solution does not exist at least for finite $x_{\rm out}$.
If the error continued to decrease with the increase in $x_{\rm out}$,  
the solution would exist for $x_{\rm out}=\infty$.
However, the fact is that the error becomes larger 
with the increase in $x_{\rm out}$
in the range $x_{\rm out}\gtrsim \log(4L)$, raising doubts for
the existence of the solution also for $x_{\rm out}=\infty$.

For small $(1/L)$, the solution exists at the
perturbative level \cite{KSSW04} and our numerical data is consistent
with this study. But we also found the nonsystematic
error that grows nonlinearly with respect to $(1/L)$.
Here, let us recall the fact that the regularity of the horizon
is not perfectly guaranteed in the perturbative study \cite{KSSW04} 
(see also Appendix A). 
If the perturbation becomes singular at the higher-order
perturbation, what happens in the numerical calculation?
Such a singular effect should appear as the numerical error
that behaves in a nonsystematic way. The nonlinear growth
of the nonsystematic error with respect to $(1/L)$
supports this view. We consider that
this is the most natural interpretation of our result.
If this is the case, the regular solution does not exist
at least to this setup.

It is interesting to ask at which order the perturbation
becomes singular. This is a difficult question to answer,
and we just make a speculation here. 
In Figs.~\ref{log10IL-log10deltaY} and 
\ref{log10IL-log10deltakappa},
we found that the nonsystematic
error grows quite rapidly, $\sim (1/L)^{4}$,
as $(1/L)$ is increased.
If we assume that this power of the nonsystematic error 
reflects the order at which the perturbation becomes singular,
the singular behavior might appear at the 4th-order perturbation.

Readers might wonder why we do not claim the existence 
of the solution at least for the range $(1/L)\ll 1$ in spite of
the consistency of our numerical data with
the perturbative study and the small values of 
the nonsystematic error.
In fact, we have reproduced the results of the perturbative study 
for $(1/L)\ll 1$ with a typical numerical error
$\lesssim 0.1\%$. However, the merits in the
numerical calculation is the ability to explore the nonlinear regime
in $(1/L)$, and our results obviously show that something is wrong
at the nonlinear level.
In such a situation, it is much more
dangerous to claim the existence of the solution
however small the detected error is. We consider that this is a common
sense for numerical researchers 
who are serious in analyzing the numerical errors.

Readers might also expect that the solution exists
for small $(1/L)$ and it vanishes at
some critical point as $(1/L)$ is increased,
as suggested in Ref.~\cite{T07}. We do not consider that this is the
case, because as shown in Figs.~\ref{xout-log10deltaY} and 
\ref{xout-log10deltakappa}, the nonsystematic
error becomes relevant for $x_{\rm out}\gtrsim \log(4L)$
for very small $(1/L)$ values such as $(1/L)=0.005$.
To support this view, 
let us look at the existing example by the author and his collaborator 
\cite{YN02} where disappearance of a solution was actually observed.
In that paper, we numerically studied the
apparent horizon formation in high-energy particle collisions,
and found that the solution of the apparent horizon exists for a small
impact parameter $b$, but it disappears at
the maximal impact parameter $b_{\rm max}$. 
As shown in Fig.~2 of Ref.~\cite{YN02}, there is a clear signal
of $b_{\rm max}$ such that $dr_{\rm min}/db$ diverges.
Although the numerical error grows as $b$ is increased, 
the error turns out to be the grid error whose growth is
driven by the distortion of the apparent horizon.
This is the typical phenomena that happens when a solution
vanishes as the system parameter is changed.
However, in the present setup, we do not find any signal
of disappearance of the solution.
Besides, the numerical error that grows with $(1/L)$
is not the grid error, but has turned out to be the nonsystematic
error. These features are all in contrast to the
case of \cite{YN02}.

Finally, let us discuss the third possibility (III)
that the solution exists but it is numerically unstable. 
Here, we refer to the case that a regular solution exists
and that two or more solutions to the finite difference equations
exist. Among these numerical solutions,
only one solution approximates the regular solution,
and our obtained solution is one of the other 
artificial solutions that do not represent the true solution.
We consider that this case is unlikely, since our solution
reproduces the result of the perturbative study for $(1/L)\ll 1$
and hence it does not seem to be artificial.
However, at the same time, we cannot rigorously show
the uniqueness of the solution to the finite difference equations 
since they are highly complicated. As a result, 
we cannot rigorously claim the nonexistence of the solution,
i.e., that the possibility (II) is the case. For this reason, our results
also have to be carefully interpreted as well as the other
numerical or perturbative studies. In order to show the nonexistence
of solutions rigorously, a proof by the method of 
the mathematical relativity will be required.

%
%======================================%
%<<<<<<<<<<<< SECTION VI  >>>>>>>>>>>>>>%
%======================================%
%
\section{Summary and discussion}
\label{Sec:VI}

In this paper, we studied a black hole on a brane
in the RS II scenario using the numerical code 
having almost 4th-order accuracy. 
Although we successfully reproduced the Schwarzschild
solution in the case of a zero tension brane (Sec.~\ref{Sec:IV}), 
the nonsystematic
error was detected for the cases of a nonzero tension brane 
(Secs.~\ref{Sec:VA} and \ref{Sec:VB}).
This nonsystematic error becomes relevant for 
$x_{\rm out}\gtrsim \log(4L)$ even for a very small $(1/L)$,
and it grows nonlinearly with respect to $(1/L)$
for a fixed $x_{\rm out}$. 
On the other hand, our numerical data agrees very well
with the perturbative study \cite{KSSW03,KSSW04} if the
numerical error is ignored (Sec.~\ref{Sec:VC}).
In Sec.~\ref{Sec:VD}, we discussed the interpretation of our result.
The most natural interpretation is that the perturbation becomes
singular at the higher order in $(1/L)$, and this effect 
appears as the nonsystematic error in the numerical data 
[the possibility (II)]. If this is the case, no solution of a
black hole on a brane exists. We also pointed out the remaining
possibility that the solution exists [the possibility (III)],
although it is unlikely.
To summarize, our numerical results raise strong doubts
for the existence of a static black hole on a brane.
To be precise, our claim is: 
{\it A solution sequence of a static black hole on 
an asymptotically flat brane
that is reduced to the Schwarzschild black hole in the zero tension limit
is unlikely to exist.}
%Our study was limited to relatively small $(1/L)$ values,
%since the numerical relaxation becomes unstable
%at some point as $(1/L)$ is increased.
%If we try to go beyond that point, the instability
%becomes more and more strong. This may indicate the nonexistence
%of a large black hole on a brane. 
%Therefore, our result may support the conjecture
%for nonexistence of a large black hole on a brane \cite{T02,EFK02}.
It is also worth pointing out that a similar claim 
has been made in the context of the
AdS/CFT correspondence \cite{FFNOS05}.

In this paper, we focused our attention to the
numerical calculations. So far, we have not succeeded
in proving the nonexistence of solutions
by an analytic method. But, here, 
we would like to give an analytic discussion that supports 
it. % and that could give a hint to prove it.
Let us ask in this way: Suppose the functions $T$, $R$,
and $C$ satisfy Eqs~\eqref{EE1}--\eqref{EE3} and \eqref{EE5}
and the boundary conditions at infinity \eqref{outer-boundary}, at
the symmetry axis \eqref{BC:symmetry1}--\eqref{BC:symmetry2}, 
on the horizon \eqref{BC:horizon1}--\eqref{BC:horizon4}, and 
${T_{,\chi}}/{T}=R_{,\chi}=({\rho}/{\ell})(e^{R}-1)$
on the brane. Then, can the function $C$ satisfy the boundary condition
$C_{,\chi}=({\rho}/{\ell})(e^{R}-1)$ on the brane?
To answer this question, we calculate the difference 
of the equations \eqref{EE3} and \eqref{EE5}.
The result is the equation of the form
%===========<Equation>============%
%
\begin{equation}
-\rho^2 C_{,\chi\chi}+C_{,\rho\rho}
+F[T_{,\chi\chi},T_{,\chi},R_{,\chi},C_{,\chi},T_{,\rho},
R_{,\rho},C_{,\rho},T,R,C,\rho,\chi]=0.
\end{equation}
%
%=================================%
Here, there is no singularity in this equation at $\chi=0$.
This is a wave equation where $\chi$ is the ``time''
coordinate. The ``initial condition'' is given by $C=R$ and $C_{,\chi}=0$
at $\chi=0$. For given $T$ and $R$, we can ``evolve'' $C$ using this equation
from $\chi=0$ to $\pi/2$. Then, the value of $C_{,\chi}$ at $\chi=\pi/2$ 
is determined as a result of the ``temporal evolution'', and therefore,
there is no guarantee that $C$ satisfies the junction condition
on the brane $C_{,\chi}=({\rho}/{\ell})(e^{R}-1)$.
Although this condition can be satisfied in a special situation
(e.g. the spherically symmetric case),
such a chance cannot be expected in general.
We consider that this discussion is consistent with our
numerical results. In Fig.~\ref{threeD-L16.6-Y}, we found an unnatural
jump in the value of $Y$ in the neighborhood of the symmetry axis.
This would be because the boundary conditions on the brane and 
at the symmetry axis are incompatible.

Suppose solutions of a static black hole on a brane
do not exist in the RS scenarios as discussed in this paper.
Then, what happens after the gravitational collapse?
As an example, let us consider a head-on collision
of high-energy particles. Because of the nonexistence
of the solution,
one might expect that the black hole does not
form in the collision. 
However, we do not consider that this is the case,
since there exist studies on the
apparent horizon formation in the RS II scenario \cite{SS00,TT07}, and
even the formation of a very large apparent horizon 
is reported \cite{TT07}. For this reason, the black hole should form.
Although the produced black hole relaxes to 
a quasi-equilibrium state by emitting the gravitational radiation,
it does not completely become static and 
cannot remain on the brane.
In this sense, the black hole is unstable. 
We suggest two possibilities
of the consequences of the instability: the event horizon pinch
and the brane pinch.

In the first possibility of the event horizon pinch, 
we refer to the case
where the brane scarcely moves and the cross section
of the event horizon and the brane shrinks. 
In the classical general relativity, the pinch of an event
horizon is forbidden (e.g. \cite{W84}).
However, if the radius of the cross section becomes order of 
the Planck length, the event horizon pinch
could happen by quantum gravity effects
and the black hole may leave the brane.
We consider that this scenario may happen
when the horizon radius is sufficiently larger than
the bulk curvature scale, since the brane is expected to be rigid
in this case.

In the second possibility of the brane pinch,
we refer to the case
where the event horizon scarcely moves while the
brane moves. The motion of the brane continues until
it pinches and generates a baby brane. We consider that
this scenario may happen when the
horizon radius is smaller than the bulk curvature scale,
since in this case the brane can easily fluctuate.
Basically, this scenario
was already suggested and studied in detail
in \cite{FT05,FPST06-2,FPST06-3,FT07}
by numerical simulations for a zero tension brane
and a more detailed model of a brane (but without
the self-gravity effects).
In those studies, the authors gave the brane an initial velocity
in the bulk direction, and observed the brane pinch. 
The new insight from the results
in the present paper 
is that such a brane pinch might happen because of
the instability without the initial velocity
in the bulk direction. We are also able to
estimate the time scale of this phenomena.
In our numerical data, the nonsystematic error
has the order of $\sim (1/L)^4$. Since this error amount is
expected to cause the temporal evolution, the typical time scale
would be $\sim \ell^4/\rho_h^3$. Therefore,
the brane pinch may happen relatively slowly
unless the velocity in the bulk direction 
is given. In order to clarify if this is the case,
a simulation taking account of the
self-gravity effects of the brane is necessary. It also has to be
checked if the brane pinch can happen when the $\mathbb{Z}_2$-symmetry
is imposed.

The above discussion has the following implication
for the black hole production at the LHC.
If the extra dimension is warped, the produced black hole
may escape into the bulk. If this is the case,
it is difficult
to observe the signals of the Hawking radiation
especially from black holes with large mass, since
the time scale of the instability may be shorter than the
time scale of the evaporation. Therefore, the instability
of black holes on a brane gives us many interesting and
important issues to be explored.

%
%======================================%
%<<<<<<<<< Acknowledgements >>>>>>>>>>>%
%======================================%
%
\acknowledgments

The author thanks the participants of the BIRS conference,
``Black Holes: Theoretical, Mathematical and Computational Aspects,''
held in Banff from November 9 to 14 (2008).
Specifically, conversations with 
D.~Gorbonos, K.~Murata, S.C.~Park, G.~Kang,
J.~Kunz, B.~Unruh, and V.P.~Frolov were fruitful.
H.Y. thanks the Killam trust for financial support.
This work was also supported in part by the JSPS Fellowships
for Research Abroad.

\appendix

%
%======================================%
%<<<<<<<<<<<< Appendix A  >>>>>>>>>>>>>>%
%======================================%
%

\section{A brief review of the perturbative study}
\label{Sec:A}

Karasik {\it et al.} \cite{KSSW03, KSSW04} 
solved the static solution of a black hole
on a brane in the RS scenarios in the case of a small mass
by a perturbative study.
Here, we summarize the part of their results that is directly related
to the discussions in this paper. In this section, we follow their notation.
Since they were mainly motivated by the TeV gravity scenarios,
they studied in the framework of the RS I scenario.
However, their results can be applied also to the RS II scenario
by changing the sign of $\ell$.

They considered the situation where the radius of the black hole
$\mu:=\sqrt{8G_5M/3\pi}$ is much smaller than the bulk curvature
scale $\ell$, and adopted the so-called matching method.
In this method, the spacetime is a perturbed Schwarzschild
black hole in the neighborhood of the horizon and is a perturbed
AdS spacetime at the distant region, where $\epsilon:=\mu/\ell$
is the small parameter. In the matching region
$\mu\ll r\ll l$, the two approximations both hold and the two
solutions can be matched to each other.

In the neighborhood of the horizon, 
the metric is assumed to be
%===========<Equation>============%
%
\begin{equation}
ds^2=\left(\frac{\ell}{z}\right)^2
\left[-B dt^2+\frac{A}{B}d\rho^2
+2Vd\rho d\psi+\rho^2 Ud\psi^2
+\rho^2\sin^2\psi d\Omega_2^2\right],
\end{equation}
%
%=================================%
where $z:=\ell-\rho\cos\psi$ and $A$, $B$, $V$, $U$ are
the functions of $\rho$ and $\psi$. 
The background solution is the Schwarzschild solution
$A_0=U_0=1$, $B_0=1-1/\rho^2$, and $V_0=0$. The first order
quantities for these functions are given as $\epsilon A_1$,
$-\epsilon B_1$, $\epsilon U_1$ and $\epsilon V_1$, respectively,
and it turns out that they are expressed in terms of 
$F$, $H$, $F_{,\rho}$, $H_{,\rho}$, $F_{,\psi}$ and $H_{,\psi}$,
where $F$ is a gauge function and $H$ is a wave function.
The wave function $H$ satisfies the equation
%===========<Equation>============%
%
\begin{equation}
(\rho^2-1)\left(H_{,\rho\rho}-\frac{H_{,\rho}}{\rho}\right)
+H_{,\psi\psi}+2\frac{\cos^2\psi+1}{\sin\psi\cos\psi}H_{,\psi}=0.
\label{waveequation}
\end{equation} 
%
%=================================%

It was found that when the perturbed AdS metric at the distant
region includes only integer powers of $M$, the matched
solution in the neighborhood of the horizon is singular
at $\rho=1$. However, if we allow the terms with half integer
powers of $M$ at the distant region, the metric becomes 
at least $C^1$ and the zeroth and first law
of the black hole thermodynamics can be satisfied.
The solution of $H$ is
%===========<Equation>============%
%
\begin{multline}
H=\delta_0-\frac{2}{3\pi}\tilde{Q}_{-1/2}(\rho)g^{(1)}(\psi)
+\tilde{Q}_{1/2}(\rho)\left(-\frac{1}{3\pi}g^{(2)}(\psi)
+\frac{d_2}{\sin^3\psi}-\frac{\pi}{3}g^{(3)}(\psi)\right)
\\
+\frac{\pi}{3}\left(\tilde{P}_{1/2}(\rho)+\frac{2}{\pi^2}
\left.\frac{\partial\tilde{Q}_{n-1/2}}{\partial n}\right|_{n=1}\right)
g^{(2)}(\psi)
+\sum_{\lambda=3}^{\infty}
a(\lambda)
\tilde{Q}_{(\lambda-1)/2}(\rho)G(\lambda,\psi).
\label{solution-H}
\end{multline}
%
%=================================%
Here, 
$\tilde{P}_{\nu}:=\rho\sqrt{\rho^2-1}P^1_{\nu}(2\rho^2-1)$,
and $\tilde{Q}_{\nu}:=\rho\sqrt{\rho^2-1}Q^1_{\nu}(2\rho^2-1)$, and
$g^{(1)}$, $g^{(2)}$, $g^{(3)}$ and $G(\lambda,\psi)$
are some elementary functions of $\psi$ (see Eq. (36)
in \cite{KSSW04} for details).
The values of $\delta_0$, $d_2$, and $a(\lambda)$ are
fixed as $\delta_0=d_2=0$ and $
a(\lambda)=-(16/3\pi)(-1)^{\lambda}/(\lambda^2-4)$
by imposing boundary conditions and the zeroth and first laws of
the black hole thermodynamics.
The expansion of the function $\tilde{Q}_\nu$ 
has the terms proportional to $(\rho^2-1)^n\log(\rho^2-1)$
for integer $n\ge 1$.
Hence, $H_{,\rho}$ has the form
%===========<Equation>============%
%
\begin{equation}
H_{,\rho}\simeq\frac{h_0(\psi)}{\sin^3\psi}
\left[\log(\rho^2-1)+2\gamma_E+2\bar{\psi}(1/2)\right]
+h_1(\psi),
\end{equation}
%
%=================================%
in the neighborhood of the horizon, 
where $h_0(\psi)$ and $h_1(\psi)$ are given by Eqs. (44)
and (45) in Ref.~\cite{KSSW04}. Here, a miraculous cancellation
happens and $h_0(\psi)$ becomes zero. This means that
the term proportional to $(\rho^2-1)\log(\rho^2-1)$
is not included in $H$. 
Then, the equation~\eqref{waveequation}
implies that $H_{,\rho\rho}$ is also finite. As a result,
the function $H$ includes terms proportional to 
$(\rho^2-1)^n\log(\rho^2-1)$ only for $n\ge 3$.

When the horizon is regular, the location of the
horizon $\rho_H$ and the surface gravity $\kappa$ are given by
%===========<Equation>============%
%
\begin{equation}
\rho_H=1+\epsilon\left[-6H(1,\pi/2)+H_{,\rho}(1,\pi/2)\right];
\end{equation}
%
%=================================%
%===========<Equation>============%
%
\begin{equation}
\kappa=1+6\epsilon H(1,\pi/2),
\end{equation}
%
%=================================%
from Eqs.~(23c) and (23d) in \cite{KSSW04},
and the 3-dimensional area of the horizon
is given by
%===========<Equation>============%
%
\begin{equation}
\frac{A_H}{A_H^{(0)}}=1-\frac92\epsilon H(1,\pi/2) 
\end{equation}
%
%=================================%
from Eqs.~(27) and (30) in \cite{KSSW04},
where $A_H^{(0)}:=2\pi^2\mu^3$. 
From these formulas, we immediately find
that the product of the surface gravity 
and the proper radius of the horizon on the brane is
$\kappa\rho_H=1+\epsilon h_1(\pi/2)$.
On the other hand, the ratio 
of the effective horizon radius to the radius on the brane becomes
$(A_H/A_H^{(0)})^{1/3}/\rho_H
=1+\epsilon\left[(9/2)H(1,\pi/2)-h_1(\pi/2)\right]$.
Here, $H(1,\pi/2)=8/9\pi$ 
from Eq.~(38) in \cite{KSSW04} and
$h_1(\pi/2)$ is numerically evaluated to be $\simeq 1.087336$.
In order to apply to the RS II scenario, we change
the sign of $\ell$ and find
%===========<Equation>============%
%
\begin{equation}
\kappa\rho_H\simeq 1-2.17467\times \frac{\sqrt{2G_5M/3\pi}}{\ell};
\end{equation}
%
%=================================%
%===========<Equation>============%
%
\begin{equation}
\frac{(A_H/A_H^{(0)})^{1/3}}{\rho_H}
\simeq 1-0.371801\times \frac{\sqrt{2G_5M/3\pi}}{\ell}.
\end{equation}
%
%=================================%
These formulas are used in Figs.~\ref{sgxrad4}
and \ref{effective_brane_radius}, respectively.

We comment on the remaining issues in this study.
In order to guarantee the regularity on the horizon, the Kretchmann scalar
$K:=R_{\mu\nu\rho\sigma}R^{\mu\nu\rho\sigma}=K^{(0)}+\epsilon K^{(1)}
+\epsilon^2K^{(2)}+\cdots$ has to be finite on the horizon.
If the function $H$ includes a term proportional to 
$(\rho^2-1)^3\log(\rho^2-1)$, the third derivative $H_{,\rho\rho\rho}$
logarithmically diverges at $\rho=1$. 
Although the authors of Ref.~\cite{KSSW04} have confirmed the absence of 
the term $H_{,\rho\rho\rho}$ in $K^{(1)}$, there remains a possibility that
$K^{(2)}$ diverges.
In order to check this, two things are required.
First, one needs to confirm if $H_{,\rho\rho\rho}$ really possesses
such a logarithmic term.
But this is  difficult because 
the coefficient of $(\rho^2-1)^3\log(\rho^2-1)$ in Eq.~\eqref{solution-H}
has the infinite summation 
of terms whose absolute values become unboundedly 
large as $\lambda$ is increased.
It turns out that this infinite summation does not converge 
even in the sense of distribution.
Therefore, there is an ambiguity in the interpretation 
of Eq.~\eqref{solution-H}.
In the particle physics, several procedures to regularize diverging infinite
summations are known. Although it may be possible to regularize 
the infinite summation in $H$ in a similar way, this is not an easy task, and
this situation makes it difficult
to confirm the (non)existence of the term proportional to 
$(\rho^2-1)^3\log(\rho^2-1)$
even numerically.
Next, the presence of a logarithmic term in $H_{,\rho\rho\rho}$ 
does not immediately imply the divergence of the Kretchmann scalar $K$,
since the terms from the 2nd-order perturbations may cancel this term.
However, the study of the 2nd-order perturbation is also a difficult task.

Finally, even if the regularity of the perturbation is established 
to some order, there remains the possibility that the metric
becomes singular at the next order.
If our interpretation in Sec.~\ref{Sec:VD} is correct, the singular behavior
might appear at the 4th-order perturbation. Therefore, it may be 
very difficult to 
check the (non)existence of solutions of a black hole on a brane
by the perturbative method.

%---------   References   ---------%

%---------   References   ---------%

\end{document}